\documentclass[amssymb,prd,aps,nofootinbib]{revtex4}
\usepackage{latexsym}
\usepackage{amsmath}
\usepackage{graphicx}
\usepackage[T1]{fontenc}
\makeindex
\newcommand{\be}{\begin{equation}}
\newcommand{\ee}{\end{equation}}
\newcommand{\bn}{\begin{eqnarray}}
\newcommand{\en}{\end{eqnarray}}
\newcommand{\bns}{\begin{eqnarray*}}
\newcommand{\ens}{\end{eqnarray*}}

\def\VE*{\vec{E}^{*}}

\def\ba{\begin{eqnarray}}
\def\ea{\end{eqnarray}}

\begin{document}
%\hfill Preprint CBPF-NF-037/03
%\onehalfspacing
\title{Noncommutative geometry induced by spin effects}
%\indent
\author{L.P. Colatto}
\email{colatto@cbpf.br, colatto@cce.ufes.br}
\affiliation{~DF, CCE, UFES, Av. Fernando Ferrari, 514, CEP 29075-910, Vit\'oria, ES, Brasil}
%\affiliation{Grupo de F\'{\i}sica Te\'orica Jos\'e Leite Lopes,
%C.P. 91933, CEP 25685-970, Petr\'opolis,  RJ, Brasil}
\author{A.L.A. Penna}
\email{andrel@fis.unb.br}
\author{W.C. Santos}
\email{wytler@fis.unb.br}
\affiliation{{\small IF, Universidade de
Bras\'\i lia, CEP 70919-970, Bras\'\i lia, DF, Brasil}}

%\pacs{12.60.-i,11.10.-z,11.15.Ex,11.30.Cp}

\begin{abstract}
\noindent
In this paper we study the nonlocal effects of noncommutative
spacetime on simple physical systems. Our main point is the
assumption that the noncommutative effects are consequences of a
background field which generates a local spin structure.  So, we
reformulate some simple electrostatic models in the presence of a
spin-deformation contribution to the geometry of the motion, and
we obtain an interesting correlation amongst the deformed area
vector, the $3D$ noncommutative effects and the usual spin vector
$\vec{S}$ given in quantum mechanics framework. Remarkably we can
observe that a spin-orbit coupling term comes to light on the
spatial sector of a potential wrote in terms of noncommutative
coordinates what indicates that bound states are particular cases in this 
procedure. Concerning to confined or bounded particles in
this noncommutative domain we verify that the kinetic energy is
modified by a deformation factor. Finally, we discuss about perspectives.

\end{abstract}
\date{\today }
%\preprint{CBPF-Notas de F\´\i sica}
\maketitle
%
% ver citações
%

\section{Introduction}
Time and again physics has taken the non-commutativity property of
some mathematical structures to produce suitable models
\cite{snyder1,yang,nekra,witten,jackiw,bahns,hewett,mocioiu,calmet,berrino,rivelles}.
It is well known that the mathematical structures related to
quantum mechanics could be interpreted as a deformation of the
classical mechanics
\cite{bayen,majid}. And this deformation reflects in a
noncommutative algebra where the classical variables are taken as
operators applied to entities that live in a special space.
Formally the quantum theory involves a series of $\hbar$ whose
coefficients are functions on the phase space, and this $\hbar$
can be interpreted as the deformation factor \cite{kont}.

In the last years, noncommutative geometry has again coming up in
physics in many different contexts. Connes, Douglas and Schwarz
had introduced noncommutative tori spaces as a possible
compactification manifold of the spacetime in their pioneering
papers \cite{connes1,connes2,connes3}. In those cases,
noncommutative geometry arises as a possible scenario for the
short-distances behavior of physical theories. In quantum field
theory the introduction of a noncommutative structure to the
spacetime coordinates, at very small length scales, can introduce
a new ultraviolet cutoff, as it was formalized by
Snyder\cite{snyder2}. It leads to new developments in quantum
electrodynamics and Yang-Mills theories in the noncommutative
variables functions versions and also appears in the framework of
the string theory\cite{strin0,strin1,strin2,strin3,strin4,strin5,strin6,strin7,strin8,strin9,strin10}. Recently,
several tests have been suggested to detect noncommutative effects
in physics
\cite{test1,test2,test3,test4,test5,test6,test7,test8,test9}. It is specially worthy to note Madore's work \cite{madore}, which formalize the mathematics of noncommutative geometry and introduced some physical applications we are going to a further and in a different point-of-view analysis in the present paper.

In this paper we purpose to analyze some aspects of the
phenomenology of spatial non-commutativity in a few simple physical
systems, in particular we are going to study the form of the potentials in noncommutative coordinates and its consequences. The central feature of this issue is to consider the
influence of the local spin structure on the geometric deformation imposed in the coordinates. In this direction we are going to verify the
consequences of the spin deformation on the content of the fields involved and
on their interactions. %For instance, we can simply notice that the Coulomb
%potential, due to the spatial non-commutativity, is modified and
%it is interesting to analyze what kind of extensions the usual
%physical world can revealed in this regime.
The first model to be
investigated is the noncommutative spacetime extended
electrostatics theory, where the
\textit{a priori} motivation of this analysis is because of the
electrostatic phenomena is an essential ingredient for the
stability of the microscopic matter. To this aim, we are going to
assume that we are working with very small contributions to the lengths on the noncommutative regime.
As a consequence, the standard electrostatics to
describe usual phenomena might be modified too. So, we are going to
reassess the Coulomb potential in the noncommutative scenario where the
self-energy of the electron is developed on a non-local spacetime
induced by the geometric deformation introduced in the model.

To a further insight in the deformation of the geometry we are
going to directly associate the noncommutative resulting vector
$\theta^{i}$ (in $3D$ dimension) with the ordinary spin vector
$S^{i}=\frac{\hbar}{2}\sigma^{i}$. As a motivation we will
consider the relation between area vector $\theta^{i}$, whose
surface encloses $\vec{S}$, to the holographic principle where the
inner hyper-volume has its properties imprinted on the border
hyper-surface \cite{hol1,hol2,hol3,hol4}. Bearing this in mind we
are going to associate the noncommutative parameter $\theta$ to
uncertainty information measurements of the spin as a deformation
on the geometry. It implies that noncommutative spacetime arises
as a kind of spin deformed spacetime. The metric measurements are
redefined and we obtain that the ordinary spin-orbit coupling
emerges naturally from a effect of first-order in $\theta$ of the
Coulomb potential in a noncommutative spacetime. We are going to
show that the spin-orbit term can be derived directly from an
arbitrary noncommutative potential. The contribution of
second-order $\theta^{2}$ effect will also be analyzed. We will
show that this term changes the kinetic energy of a particle
confined on a noncommutative space, and that deformation factor of
the kinetic energy assumes a standard factor related to the ground
state of energy $E_{g}$ of the system. We are going to show that
such a phenomenon is related directly to symmetry Lorentz
violation.

This paper is outlined as follows: in section II, we present the
basic concepts of the noncommutative spacetime. In section III, we
analyze the behavior of basic electrostatics and their
fundamentals in a deformed geometry due to non-commuting space
coordinates. In section IV, we present the relation between area
vector $\vec{\theta}$ (noncommutative parameter) and spin
$\vec{S}$, and as consequence we obtain the general spin-orbit
term. In section V, we study the non-linear $\theta^{2}$ effects
on the kinetic energy of simple physical systems due to spin
deformation of spacetime. In this section, we still obtain a
standard factor of deformation for kinetic energy in
noncommutative geometry. In section VI, we present basic elements
of quantization and to suggest an interesting method to obtain the
ground state of energy $E_{g}$ of physical systems via the
standard factor of deformation of kinetic energy. In section VII,
we present the general conclusion.

\section{The non-commuting space}

In quantum mechanics the phase space can be defined replacing the
canonical variables: position $x^{i}$ and momentum $p^{i}$, by
their counterparts: the Hermitian operators $\hat{x}^{i}$ and
$\hat{p}^{i}$ \cite{madore}. These operators obey the Heisenberg commutation
relation,
\begin{equation}
\label{comm0}
\big [\hat{x}^{i}\,,\,\hat{p}_{j}\big]=i\hbar\,\delta^{i}_{j}\,\,.
\end{equation}

In such a class of theories one can easily infer about the
possible failure of the commutation property on the position
operator measurements, and this fact could be reassessed by
proposing that the spacetime coordinates operators do obey the
follow commutation relation:
\begin{equation}
\label{comm1}
\big [\hat{x}^{i}\,,\,\hat{x}^{j}\big]=i\theta^{ij}\,\,,
\end{equation}
where the parameter $\theta^{ij}$ is an  antisymmetric and
constant tensor with dimension equals to $(length)^{2}$. An
important aspect concerning this proposition is that the notion of
a ``classical point'' has no meaning any longer in the
noncommutative space, and the spacetime manifold is replaced by a
Hilbert space furnished by states which obey an uncertainty
relation as,
\begin{equation}
\label{comm2}
\Delta x^{i}\Delta x^{j}=\frac{1}{2}\left|\theta^{ij}\right|\,\,,
\end{equation}
that has the same form as the Heisenberg uncertainty principle. In
this way, a spacetime point is replaced by a Planck cell with
dimension of area.

To build a noncommutative version of a model we might to replace
the ordinary product between functions by its noncommutative
counterpart which is based on the Groenewald-Moyal
\cite{moyal,szabo} product, or star product $(*)$. This new
product operation is applied to functions of noncommutative
variables, or
\begin{equation}
f(\hat{x})g(\hat{x})\longrightarrow  f(x)*g(x) =
exp\left(\frac{i}{2}\,
\theta^{\mu\nu}\partial_{\mu}\partial_\nu\right)f(x)g(x')\big\lvert_{x^{'}=x}\,
\, ,
\end{equation}
where the star product between the functions is written as a
particular operation on functions depending on usual commuting
coordinates. In this algebra the ordinary commutator between
spacetime coordinates can be replaced by a nontrivial form given
in the expression  (\ref{comm1}). The relation between the
noncommutative variables functions and the usual ones is expressed
by
\begin{equation}
f(x)*g(x)=f(x)g(x)+
\frac{i}{2}\,\theta^{ij}\partial_{i}f(x)\partial_{j}g(x)+{\cal O}(\theta^{2})\, ,
\end{equation}
where the representation of the product $f(x)*g(x)$ indicates a
deformed algebra of functions on a general space $\mathbb{R}^{3}$
or a noncommutative algebra. Hence such a deformation must be
connected to a noncommutative geometry by means of a Lie algebra
for coordinates $x^{i}$ on $\mathbb{R}^{3}$ represented by the
equation (\ref{comm1}). Remarkably, it is possible to connect the
noncommutative algebra (\ref{comm1}) to the Heisenberg uncertainty
relation (\ref{comm0}) considering the parametric noncommutative
spacetime coordinate,
\begin{equation}
\label{commp}
\hat{x}_{i}=x_{i}-\frac{\theta_{ij}p_{j}}{2\hbar}.
\end{equation}
We can easily notice that this relation satisfies the algebras
(\ref{comm0}) and (\ref{comm1}) and it also exhibits a
non-locality feature of the theory in a particular simple way.
Furthermore it gives rise to a principle which says that for a
large momenta we have a large non-locality. This non-locality can
be depicted observing that a plane wave corresponds no longer a
point particle, as in commutative quantum field theory, but
instead a ``dipole''. Indeed it refers to a rigid oriented rod
which the extension is proportional to its momentum $\Delta
x_{i}=\theta_{ij}p^{j}/2\hbar$. In this case, we can propose a
general postulate in order that such ``dipoles'' interact amongst
themselves sticking their ends, similar to the open strings
\cite{nekra}.

\section{Electrostatics in a noncommutative spacetime}

To analyze the electrostatic case we begin our investigation
considering a simple model describing electric charges living on a
noncommutative geometry. As a prototype application we will
introduce briefly only some basic rules to electrostatics in a
spatial noncommutative scenario. To this aim, we define the
distance $\hat{r}$ in this space which undergoes the influence of
the deformed geometry. The simplest way to perform this issue is
to consider that a modified electric force is a consequence of
simply changing the commutative coordinates to the noncommutative
ones and implement the ``new'' algebra coordinates. So we take the
noncommutative electric potential $\widehat{V}(\hat{r})$ as an
extension of the usual electric potential in a noncommutative
spacetime, or it now depends on the noncommutative position
$\hat{r}$, and on the usual electric charge $q$.  The
noncommutative distance $\hat{r}$ we can define by means of the
usual inner product onto noncommutative coordinates $\hat{x}^{i}$,
in the follow form,
\begin{equation}
\hat{r}^{2}=<\hat{x}^{i},\hat{x}^{i}> \,,
\end{equation}
which measures the ``deformed'' line length, or distance. Using
the algebraic relation (\ref{commp}) we find the general
expression,
\begin{equation}
\label{rcomut}
\hat{r}^{2}=r^2 + \rho^2=r^2+\frac{\vec{\theta}\cdot\vec{L}}{\hbar}+
\frac{\big|\,\vec{\theta}\times \vec{P}\big|^{2}}{4\hbar^{2}}\,,
\end{equation}
where $r$ is the ordinary distance, and $\rho$ is the radius of
deformation of the space which is independent of $r$ in $3D$. Then
if $\rho\neq 0$ we have a nonlocal spacetime and the classical
geometry arises deformed. If $\rho=0$ we reach the local spacetime
regime. We adopt also, by simplicity, $\vec{L}$ constant in time
and position. In $3D$ the vector $\vec{\theta}$ can be the dual of
the deformation tensor (or noncommutative tensor) $\theta_{ij}$,
and furthermore can represent an arbitrary noncommutative
vector parameter which can be written as %$\vec{\theta}=\theta_i=\big(\theta_{1},\theta_{2},\theta_{3}\big),$
\begin{equation}
\label{tcomut}
\vec{\theta}=\theta_i=\big(\theta_{1},\theta_{2},\theta_{3}\big)=-\varepsilon_{ijk} \theta_{jk},
\end{equation}
where the $\theta_{i}$ components can assume any real value. The
constant vector $\vec{\theta}$ represents an uncertainty parameter
to the simultaneous space coordinates measurements. In this
particular choice, the noncommutative algebra takes the form of a
Lie algebra, therefore we are going to have a kind of rotation
symmetry in the coordinates, and consequently it should be
correlated to the spin structure. Moreover, we can observe that
the presence of the linear momentum $\vec{P}$ and angular momentum
$\vec{L}$ in the expression \eqref{rcomut} is a consequence of
this deformation of the spacetime. In this sense it is possible to
perform a mapping from a general and well-defined noncommutative
function $\widehat{f}(\hat{r})$ to a deformed function
$f\big(r\,,\,\rho)$ on noncommutative geometry,
\begin{equation}
\widehat{f} \big(\hat{r}\big) \longmapsto  f\big(r\,,\,\rho).
\end{equation}
From the deformed distance (\ref{rcomut}) we emphasize that  the
deformed radius $\rho$ can
 be written as,
\begin{equation}
\rho^{2}=\frac{\vec{\theta}\cdot\vec{L}}{\hbar}+
\frac{\big|\,\vec{\theta}\times
\vec{P}\big|^{2}}{4\hbar^{2}},
\end{equation}
where we include all the deformed terms which refer to the linear
momentum $\vec{P}$ and the angular momentum $\vec{L}$. Then $\rho$
denotes the radius of an elementary volume of the spacetime for
small distances measurements. We observe that the expression of
noncommutative distance (\ref{rcomut}) is the most general
deformation which satisfies the noncommutative algebra
(\ref{comm1}), in fact we shall still stick out that no
deformation is attributed to $\theta^{3}$ term that satisfies the
relation (\ref{comm1}) and (\ref{commp}).  An important point to
remark is that noncommutative spacetime models and their
corresponding deformations induces a nonlocal spacetime
configuration, and this non-locality property implies to a
non-conventional charge distribution in this noncommutative
regime. Then, considering this fact, the notion of point charge
turn to be not suitable when we treat these effects at very small
length.

Now, to build the potential energy \footnote{We are adopting
gaussian units.} term we assume that two charge $q$ and $q'$ are
separated by a noncommutative distance $\hat{r}$, so we can write
down
\begin{equation}
\label{encomut}
\widehat{V}=\frac{qq'}{\hat{r}}=\frac{qq'}{\sqrt{r^{2}+\rho^{2}}}.
\end{equation}
Then bearing in mind that the noncommutative and nonlocal effects
are consequences of the deformation $\rho$, the noncommutative
electric potential $\widehat{\Phi}(\hat{r})$ (due a source charge
$q$) in a noncommutative geometry can be written as,
\begin{equation}
\label{vcomut}
\widehat{\Phi}(\hat{r})=\frac{q}{\hat{r}}=
\frac{q}{\sqrt{r^{2}+\rho^{2}}}\,.
\end{equation}
The norms of electric field $\widehat{E}(\hat{r}) $ and the
Lorentz force $\widehat{F}(\hat{r})$, are also written as
\begin{eqnarray}
\label{ecomutt}
\widehat{E}(\hat{r})=\big|\vec{\nabla} \widehat{\Phi}(\hat{r})\big|
=\frac{qr}{(r^{2}+\rho^{2})^{\frac{3}{2}}}\; \; \;
\mbox{and } \;\;
\widehat{F}(\hat{r})=q^{\prime}\frac{\partial \widehat{\Phi}(\hat{r})}{\partial
r}=\frac{qq{^\prime}r}{(r^{2}+\rho^{2})^{\frac{3}{2}}}\,.
\end{eqnarray}
The figure below shows that the electrostatic force between two
elementary charges $e$ (in vacuum) decreases for very small length
scales ($\approx 100fm$) when we assume to deal with
noncommutative geometry. This behavior implies obviously in a
non-conventional electric force theory. Moreover, we can note that
for distances at atomic order ($>1pm$) the model reaches the
well-known conventional Coulomb force. We also observe that the
modified electric force fades smoothly up to zero in the Planck
length ($\approx 10^{-33}cm$).
\begin{center}
\begin{figure}[h]
\begin{centering}
\rotatebox{0}{\resizebox{9.5cm}{!}{\includegraphics{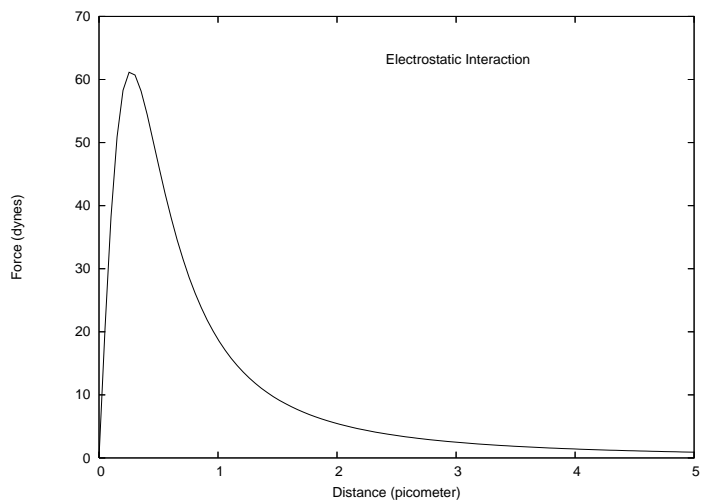}}}
\caption{Plot of electrostatics force $\widehat{F}$ versus
the distance $r$.}
\label{none}
 \end{centering}
\end{figure}
\end{center}
It is easy to verify that at the limit $\rho\rightarrow 0$ we also
reach the commutative regime and consequently the model shows an
ordinary electrostatics regime at this limit.

Assuming that $\rho$ is small in comparison with $r$ we can expand
the the scalar potential \eqref{vcomut}, and taking the terms up
to the order $\vec{\theta}^2$ we have,
\begin{equation}
\label{ecomut}
\widehat{\Phi}(\hat{r})\simeq \frac{q}{r}-
\frac{q}{2\hbar}\frac{\vec{\theta}\cdot\vec{L}}{r^{3}}
-\frac{q}{8\hbar^{2}}\frac{\big|\,\vec{\theta}\times
\vec{P}\big|^{2}}{r^{3}}+\frac{3q}{8\hbar^{2}}\frac{\big(\vec{\theta}\cdot\vec{L}\big)^{2}}{r^{5}},
\end{equation}
where the first term $q/r$ stands for the ordinary electric
potential $\Phi(r)$ in the stationary regime. The remained terms
in (\ref{ecomut}) indicate the typical structure of deformations
up to the chosen order. In fact the expression (\ref{ecomut}) is
equivalent to standard Moyal noncommutative expansion of the
function $\widehat{\Phi}(\hat{r})$ \cite{moyal} which is given by,
\begin{equation}
\label{ecomut2}
\widehat{\Phi}(\hat{r})= \Phi(r)-
\frac{1}{2\hbar}\theta_{ij}P^{j}\partial^{i}\Phi(r)+
\frac{1}{8\hbar^2}\theta_{ij}\theta_{kl}P^{j}P^{l}\partial^{i}\partial^{k}\Phi(r)+...+{\cal
O}(\theta^{n}),
\end{equation}
where we can easily  verify that the expressions \eqref{ecomut}
and \eqref{ecomut2} are equivalent, and moreover it is again easy
to verify from the expression \eqref{ecomutt} that
$\vec{\nabla}\times\widehat{E}(\hat{r})=0$. On the other hand, the
notion of point charge is no longer sensible, due to the
``nonlocal'' effects present in this model which implies that
$\vec{\nabla}\cdot\widehat{E}(\hat{r})\neq0$ and to the
noncommutative $\vec{\theta}$ parameter. Remarkably at very small
distances limit $\vec{\nabla}\cdot\widehat{E}$ converges
 to a fix value associated to elementary area describes by $\theta$, or
\begin{equation}
\label{limE}
\lim_{r\rightarrow 0}\vec{\nabla}\cdot\widehat{E}=4\pi\frac{e}{\Omega},
\end{equation}
where $e$ is the elementary charge distributed over the
deformation volume ${\Omega}=4\pi
\rho^{3}/3$. So in this limit we can infer that the ordinary spacetime configuration looses its classical form and may be redefined incorporating new structural
elements beyond the geometrical notion of point. In our study,
such elements could be regarded as a kind of micro-deformations of
the spacetime, which are not connected to usual idea of topology
of the space, but associated to possible (quantum) dynamics of the
deformations. Of course these non-trivial topological effects are
originated from $\rho$ deformations of space and if we take the
usual topology, this deformation could be correlated to anomalies
that appear in a non-conventional discrete spacetime. On the other
hand, for large distances or $r^{2}>>\rho^{2}$ the conventional
electrostatics effects prevails over the noncommutative ones.

The scope of the present work is not to enter in the mathematical
formalism details, instead our aim is to present and discuss some
interesting new features to the study of electrostatics in the
noncommutative space extension. One of these features is the
self-energy of the charged particle, which is taken as an
important property induced by the deformation radius $\rho$. It is
well-known in the usual classical electrostatics the self-energy
of a point charge diverges while it is easy to infer from the
expression \eqref{limE} that, in a noncommutative space, the
electrostatics self-energy $\widehat{U}$ becomes ``non-punctual''
and the expression can be written as
\begin{equation}
\label{self}
\widehat{U}=\frac{1}{2}\int^{\infty}_0 \big|\widehat{E}\big|^{2}d^{3}x
=\frac{3\pi^{2}e^2}{8\rho},
\end{equation}
where we have used the expression (\ref{ecomutt}), and we can
notice that at the infinity limit the energy goes to zero as in
the usual case, but at the zero limit we obtain an exact and
finite
 value,
%\begin{equation}
%\widehat{U}=\frac{3e^2}{128\epsilon_{0}\rho},
%\end{equation}
which is a function of the radius of deformation $\rho$ only.
Therefore, we can also conjecture that the elementary electric
charge $e$ can come up enclosed into a minimal surface which is
characterized by an intrinsic radius of curvature of the
deformation $\rho$. Hence this nonsingular convergence of the
equation $\widehat{U}$ that could be an effect of the deformation
on the spacetime geometry results in a nonlocal effect
\cite{ahwalia1,ahwalia2} when we consider very small lengths.
Then we can conjecture that we have obtained clue of a possible
internal structure of the charged particle depicted by deformation
radius $\rho$ of the spacetime.

\section{Noncommutative spacetime and Spin}
\noindent

An important issue to study is the possible origin of the
deformation radius $\rho$. In the case treated here, for a very
small length, it is reasonable to evaluate that the continuous
structure of the spacetime might be spoiled and, consequently,
nonlocal effects emerge. Recently, the discrete spacetime
structure resulted \cite{rovelli1} (or atomic-like structure of
spacetime) has been object of intense study where nonlocal effects
are correlated to a discrete spacetime structure via a spin foam
hypothesis
\cite{foam1,foam2,foam3,foam4}, where noncommutative geometry emerges
naturally. In our case we are going to simply explore the
noncommutative algebra
\eqref{comm1}, where we introduce further properties associated to the noncommutative
antisymmetric tensor $\theta^{ij}$ in order to obtain models that
could be phenomenological. Indeed in $2D$ it is usual to directly
associate the tensor $\theta^{ij}$ to the Levi-Civita tensor
$\epsilon^{ij}$. In that case it is possible to build various
theoretical models with a simple symplectic structure. On the
other hand, in $3D$ we can directly connect the tensor
$\theta^{ij}$ to the 3-index Levi-Civita contracted to the dual
vector, or $\epsilon^{ijk}\theta_k$.  In this point view, we might
also conjecture that the measurements of $\theta$ implies in
nonlocal effects due to possible discrete spacetime structure. On
the other hand, let be $\theta$ the area of a connected
two-dimensional spatial surface (or event horizon) that contains
the spin $\vec{S}$ and guard all information of the intrinsic
angular momentum. So we can consider the spin $\vec{S}$
proportional to area $\theta$ of this surface. We have reasons to
believe that this concept is associated to the holographic
principle
\cite{hol2,hol3}. Then, for very small lengths, we can conjecture that
this event horizon can modify the classical geometry structure by
the spin.

We must to do some comments. In quantum mechanics the realization
of the spin as an observable is based on the application of an
external magnetic field. This implies that for the spin be
``realized'' it is necessary to embed the system into an
environment fulfilled with a magnetic field. In fact, it is
possible to verify from the literature in some cases that in a
noncommutative theory we can associate the $\theta$ parameter to
the magnetic field $B$, or $\theta\propto 1/B$
\cite{nekra,szabo,barbon}. In this sense we propose to deform the
geometry ``seeing'' by an observer on a charged particle and
consequently the particle Lorentz symmetry is violated
\cite{viol1,viol2,viol3,viol4,viol5}. So the curved trajectory
done by the charged particle reflected by the a ``spin'' vector in
the noncommutative algebra \eqref{comm1} indicates that it is
related to that background field environment and a Lorentz
symmetry violation on the usual framework. Lorentz symmetry
violation must occurs from incomplete information (due to
event-horizon-like $\theta$) of spin $\vec{S}$ and the observer
(magnetic field) both limited to uncertainty and non-locality of
spacetime. In any situation it indicates that the spin deforms the
spacetime structure proportional to event horizon area $\theta$ \cite{madore}.

The relation between $\theta$ and the spin of the particle can be
shown by means of a simple and direct mathematical formalism in
$3D$ which does bring to light well-known effects. Bearing in mind
the results of last sections, we intend to fix the arbitrary
character of noncommutative area vector $\vec{\theta}$ assuming a
direct connection to the spin, or
\begin{equation}
\label{rspin}
\vec{\theta}=\frac{\hbar}{m^{2}c^{2}}\vec{S}\,,
\end{equation}
where $\vec{\theta}$ is the dual vector already given in the
equation (\ref{tcomut}) and $\vec{S}$ can be assumed as the vector
spin operator or
%\begin{equation}
$\vec{S}=(\hbar/2)\vec{\sigma}$,
%\end{equation}
where $\vec{\sigma}$ are the Pauli matrices. We must observe that
the expression (\ref{rspin}) is a relation in the quantum domain:
for very large $m$ (classical mechanics limit) so
$\theta_{i}\approx 0$ and, consequently, the classical geometry
prevails. Hence in our proposal we suggest that the noncommutative
deformation arises from the spin structure of a particular model.
Then
%\begin{equation}
%\label{r2comut}
%\hat{r}^{2}=r^2+\frac{\vec{S}\cdot\vec{L}}{m^{2}c^{2}}+
%\frac{\hbar^{2}}{16m^{4}c^{4}}\big|\,\vec{\sigma}\times
%\vec{P}\big|^{2}\,.
%\end{equation}
using the Clifford algebra
$\{\sigma_{i}\,,\,\sigma_{j}\}=2\delta_{ij}$, it is easy to get
that $\big|\,\vec{\sigma}\times
\vec{P}\big|^{2}=2P^{2}$ and, so the inner product (\ref{rcomut}), or the deformed distance,
can be rewritten as,
\begin{equation}
\label{r3comut}
\hat{r}^{2}=r^2+\frac{\vec{S}\cdot\vec{L}}{m^{2}c^{2}}+
\frac{\hbar^{2}P^{2}}{8m^{4}c^{4}}\,.
\end{equation}
We can observe that the term $\vec{S}\cdot\vec{L}$ is the
spin-orbit coupling contribution to the magnitude of the distance,
while the last term can be associated to fluctuations of the
kinetic energy.  In fact the deformation $\theta_{ij}$ of the
space can play the same role of spin influenced by an external
magnetic field involved in the electron orbital motion as in the
atomic physics. It is remarkable that the expression of the
deformed distance showed in the expression \eqref{r3comut} is the
most general extension which satisfies the noncommutative algebra
(\ref{comm1}). So the algebra can be can rewritten as,
\begin{equation}
\label{ncomm1}
\big
[\hat{x}_{i}\,,\,\hat{x}_{j}\big]=-\frac{i\hbar}{m^{2}c^{2}}\epsilon_{ijk}S_{k}\,\,.
\end{equation}
In this scenario the position operator has an Heisenberg's
uncertainty principle which depends on the background field
applied. So, in this sense, the spin vector observation is a
consequence of the presence of this field. To get a deep insight
we are going to do some applications.

\subsection{Coulomb Potential}

Recalling the Hamiltonian $H$ in ordinary spacetime with Coulomb
potential and a perturbation term for the fine structure which
contains the spin-orbit coupling
$V_{so}=\frac{e^{2}}{2m^{2}c^{2}r^{3}}\,\vec{S}\cdot\vec{L}.$
(which is used for the hydrogen atom ).
%\begin{equation}
%\label{ospin}
%$H_{so}=\frac{e^{2}}{2m^{2}c^{2}r^{3}}\,\vec{S}\cdot\vec{L}.$
%\end{equation}
The full Hamiltonian is then given by,
\begin{equation}
\label{hamil1}
H=T+V=\frac{P^{2}}{2m}+V=\frac{P^{2}}{2m}+V_{em}+V_{so}=\frac{P^{2}}{2m}
-\frac{e^{2}}{r}+ V_{so} =\frac{P^{2}}{2m}-\frac{e^{2}}{r}+
\frac{e^{2}}{2m^{2}c^{2}r^{3}} \, \vec{S}\cdot\vec{L},
\end{equation}
where $m$ denotes the reduced mass of the system. The above
Hamiltonian describes the hydrogen atom in standard quantum
mechanics. Now, taking the relations (\ref{rspin}) and
(\ref{r3comut}) we are going to plug in the noncommutative
distance and consequently the new Hamiltonian $\widehat{H}$ which
brings a noncommutative Coulomb potential as suggested by
expression (\ref{encomut}). The Hamiltonian without the spin
sector can be written as,
\begin{equation}
\label{hamil11}
\widehat{H}_{em}=T+\widehat{V}_{em} =\frac{P^{2}}{2m}-\frac{e^{2}}{\sqrt{r^2+\rho^{2}}}\,.
\end{equation}
Using the expression \eqref{ecomut2} we are going to expand the
noncommutative potential energy $\widehat{V}_{so}(\hat{r})$ up to
the linear term in $\vec{\theta}$ for simplicity, so we have, that
\begin{equation}
\label{o2spin}
\widehat{V}_{em}(\hat{r})\approx-\frac{e^{2}}{r}+
\frac{e^{2}}{2r^{3}\hbar}\,\vec{\theta}\cdot\vec{L}=-\frac{e^{2}}{r}+
\frac{e^{2}}{2m^{2}c^{2}r^{3}}\,\vec{S}\cdot\vec{L}\,,
\end{equation}
where we use the spin correlation ansatz (\ref{rspin}). It is easy
to see that the classical spin-orbit potential $V_{so}$ coincides
with the last term of the noncommutative potential (\ref{o2spin})
which is the linear term of the Taylor expansion. In
phenomenological point view, the analysis of the hydrogen atom
spectrum in noncommutative QED might suggest also the spin-orbit
effect, as was introduced by Chaichian, Sheikh-Jabbari and Tureanu
\cite{hydrogen1,hydrogen2}. We claim that our choice in order to
associate the spin $\vec{S}$ to the $\vec{\theta}$ can show that
the deformation of the space is possible due to the spin
originated from a background field as the noncommutative effect.
An interesting analogy could be made here amongst the spin
dynamics and vortice-like dynamics derived from the quantization
of the Eulerian dynamics of point vortices in an ideal fluid. In
this case, it is known that the geometrical center of coordinates
of the vortex do not commute. This similar correspondence of these
properties might suggest a relation between vortex and intrinsic
spin dynamics which could be dictated by the noncommutative
spacetime phenomenology.

\subsection{The Harmonic Oscillator}

We can extend the equivalence between noncommutative potential
energy and spin-orbit effects. To this aim we start off from the
general spin-orbit coupling of standard quantum mechanics written
in the form,
\begin{equation}
\label{o3spin}
V_{gso}=\frac{1}{2m^{2}c^{2}}\left(\frac{1}{r}\frac{dV(r)}{dr}\right)\vec{S}\cdot\vec{L},
\end{equation}
where $V(r)$ is any ordinary potential energy. As a second example
we can assume the potential energy for the simple harmonic
oscillator $V(r)=\frac{kr^{2}}{2}$, and it results that expression
(\ref{o3spin}) becomes as,
\begin{equation}
\label{o4spin}
V_{hoso}=\frac{k}{2m^{2}c^{2}}\,\vec{S}\cdot\vec{L}\,,
\end{equation}
where $k$ is elastic constant. Remarkably we can derive the
spin-orbit expression (\ref{o4spin}) from the noncommutative
potential energy taking the deformed distance and its expansion to
the linear term in $\theta$, or
\begin{equation}
\label{o5spin}
\widehat{V}_{ho}(\hat{r})=\frac{k\hat{r}^{2}}{2}\approx\frac{kr^{2}}{2}+
\frac{k}{2\hbar}\,\vec{\theta}\cdot\vec{L} = \frac{kr^{2}}{2}+ \frac{k}{2m^{2}c^{2}}\,\vec{S}\cdot\vec{L}\,,
\end{equation}
where we assume the spin correlation (\ref{rspin}) so we obtain
that the spin-orbit appears when we assume to be in a
noncommutative framework.

\subsection{Logarithmic Potential}

As a third example we can assume the logarithmic potential energy
$V(r)=V_{0}\ln\big|\alpha r\big|$. From the general spin-orbit
equation (\ref{o3spin}) we obtain that,
\begin{equation}
\label{o6spin}
V_{so}=\frac{V_{0}}{2m^{2}c\,r^{2}}\,\vec{S}\cdot\vec{L}\,.
\end{equation}
And so we again can derive this term through expansion of a
noncommutative logarithm potential energy which is function of
$\hat{r}$, resulting in,
\begin{equation}
\label{o7spin}
\widehat{V}(\hat{r})=V_{0}\ln\big|\alpha\hat{r}\big|\approx
V_{0}\ln\big|\alpha
r\big|+\frac{V_{0}}{2m^{2}c\,r^{2}}\,\vec{S}\cdot\vec{L}\,
\end{equation}
where we can see as the previous case that the spin-orbit term
emerges as a contribution of the noncommutative expansion of the
potential energy around the commutative distance.

\subsection{Yukawa Potential}
To the Yukawa potential energy we repeat the same procedure, so
taking the potential energy  $V(r)=\frac{q^{2}}{r}e^{-\alpha r}$
and using the general form of the spin-orbit coupling equation
(\ref{o3spin}) the potential becomes,
\begin{equation}
V_{so}=-\frac{q^{2}\,e^{-\alpha
r}}{2m^{2}c^{2}\,r^{3}}\big(1+\alpha r\big)\vec{S}\cdot\vec{L}\,.
\end{equation}
Which we extend to the noncommutative scenario and so we can show
explicitly the expansion of the noncommutative Yukawa potential
energy, which yields to,
\begin{equation}
\widehat{V}(\hat{r})\approx V(r)
-\frac{q^{2}\,e^{-\alpha r}}{2m^{2}c^{2}\,r^{3}}\big(1+\alpha
r\big)\vec{S}\cdot\vec{L}\,.
\end{equation}
We can conclude that for any well-defined spatial noncommutative
potential energy $\widehat{V}(\hat{r})$ its Taylor expansion from
the commutative distance around its noncommutative background
field contribution results in the ordinary potential energy $V(r)$
and an additional linear contribution of the spin-orbit coupling
term. As a matter of fact this is only valid if we assume that the
spin is correlated with the noncommutative tensor $\theta^{ij}$
showed in the expression (\ref{rspin}).

\subsection{General Case}

We can infer from a general case of the linear contribution of the
expansion of the potential energy that the correlation between
spin-orbit effects and a noncommutative potential energy
$\widehat{V}(\hat{r})$ can be directly obtained taking its
expression, or
\begin{equation}
\label{pot1}
\widehat{V}(\hat{r})=
\widehat{V}\left(\left[r^2+\frac{\vec{\theta}\cdot\vec{L}}{\hbar}\right]^{\frac{1}{2}}\right) \simeq V(r)+\frac{\vec{S}\cdot\vec{L}}{2m^{2}c^{2}}\left(\frac{1}{r}\frac{dV}{dr}\right)
\end{equation}
where we perform the expansion up to the first order around
$\vec{\theta}\cdot\vec{L}/\hbar$ and we assume the spin
correlation (\ref{rspin}). It remarkably shows that any
noncommutative potential energy can be written (as a good
approximation) as the ordinary potential energy $V(r)$ and a
general spin-orbit coupling term as the equation (\ref{o3spin}).
It suggests that noncommutative deformations could be seen as
effective quantum deformations in the spacetime constrained by the
dynamics of the spin structure.

\section{Noncommutative effects on the Kinetic Energy}
\noindent
Up to now we have only discussed the spin-orbit effects derived
from noncommutative geometry.  Now on we shall analyze the
effective contribution of the extra kinetic term $\hbar^{2}
P^2/8m^{4}c^{4}$ obtained in expression (\ref{r3comut}) for
Coulomb potential-like in Hamiltonian method. It is well-known
that for large linear momenta $P$ this term in the main one,
particularly in high energy physics. Then using the expression
(\ref{r3comut}) we can obtain the expansion of the noncommutative
Hamiltonian (\ref{hamil11}), which results in,
\begin{equation}
\label{hamil2}
\widehat{H}=\frac{P^{2}}{2m}-\frac{e^{2}}{r}+
\frac{e^{2}}{2m^{2}c^{2}r^{3}}\,\vec{S}\cdot\vec{L}+
\frac{e^{2}\hbar^{2}P^{2}}{16\,m^{4}c^{4}\,r^{3}}\,,
\end{equation}
as we have already seen the third term is the spin-orbit coupling
one, further the last term, which also decays with $r^{3}$
distance, represents a kinetic term contribution from the
noncommutative deformation. We can verify for the
expression(\ref{hamil2}) that the extra kinetic term can be
incorporates to the first term in such a way that the deformed
Hamiltonian can be written as,
\begin{equation}
\label{hamil3}
\widehat{H}=\frac{\alpha(r)P^{2}}{2m}+\widehat{V}_{em}(\hat{r})\,,
\end{equation}
with the coefficient $\alpha(r)$ is
\begin{eqnarray}
\label{fac1}
    \alpha(r)=1+\frac{e^{2}\hbar^{2}}{8m^{3}c^{4}r^{3}}
\end{eqnarray}
where it can be interpreted as a deformation factor to the kinetic
energy. According to the Hamiltonian (\ref{hamil2}) the spin-orbit
coupling appears as a noncommutative effect in potential energy
while the factor $\alpha(r)$ denotes the noncommutative influence
on the kinetic energy part. We notice that the maximal
contribution of the $\alpha(r)$ occurs when the system assumes the
ground state energy. This means that the distance $r$ is the Bohr
radius $a_{0}=\frac{\hbar^{2}}{me^{2}}$. In this scenario the
factor $\alpha(a_{0})$ is maximal and comes up dependent of the
ground state of energy $E_{g}$ as well as the  particle rest
energy $E_{r}$ of system. Hence the expression (\ref{fac1})
assumes the following form,
\begin{equation}
\label{fac2}
\alpha(a_{0})=1+\frac{1}{2}\left(\frac{E_{g}}{E_{r}}\right)^{2}\,.
\end{equation}
So to the hydrogen atom the ground state of the energy is given by
$e^{2}/2a_{0}$ while the classical rest energy is $mc^2$. For a
state with a fixed energy parameter we can determine locally the
factor indicated in (\ref{fac2}) and the kinetic energy of the
Hamiltonian becomes dependent on the rate $E_{g}/E_{r}$ of the
system. Indeed it is a consequence of our noncommutative spacetime
assumption the imposition of a self-interaction point of view. We
can also note that if $\frac{E_{g}}{E_{r}}<<1$ the factor
(\ref{fac2}) results to be $\alpha\approx 1$ and we turn back to
the classical theory. In phenomenological situations where $E_{g}$
is large we have $\alpha>1$ and it results in a interesting
noncommutative effect on the kinetic energy of system. On the
other hand to hydrogen atom we estimate that
$\frac{E_{g}}{E_{r}}\approx 7.0\times10^{-10}$  what implies  to
$\alpha\approx 1$. This result agrees to that of the classical
Hamiltonian of the hydrogen atom where (at low energy) the kinetic
term becomes unmodified. Therefore in high energy system the rate
$\frac{E_{g}}{E_{r}}$ induces an increment in the kinetic energy
which can be related to the violation of Lorentz symmetry. In fact
noncommutative theories manifest naturally Lorentz symmetry
violation as it has been shown in several article
\cite{nekra,szabo,viol1,viol2,viol3}. Such features can also be connected to the
deformation factor $\alpha$ in the Hamiltonian (\ref{hamil3}). It
is known that a typical spin-orbit effect violates symmetries of
the classical Hamiltonian in standard quantum mechanics. However,
as we have seen both the factor $\alpha$ and spin-orbit coupling
can be derived from  noncommutative effects of potential
$\widehat{V}(\hat{r})$ with manifest Lorentz symmetry violation.

Bearing this in mind we shall analyze the behavior of the harmonic
oscillator Hamiltonian (for confined fermions) assuming the
complete definition (\ref{r3comut}). Starting off from the
noncommutative harmonic oscillator Hamiltonian, where we simply
changed the ordinary distance to the deformed one, we can write
down as
\begin{equation}
\label{hamil14}
H_0= \frac{P^2}{2m}+\frac{m\omega^{2} r}{2} \longrightarrow
\widehat{H}=
\frac{P^2}{2m}+\frac{m\omega^{2} \hat{r}}{2}\,,
\end{equation}
and taking the Taylor expansion in $\hat{r}$, we find that,
\begin{eqnarray}
\label{hamil15}
\widehat{H}&=&\frac{P^2}{2m}+\frac{m\omega^{2}r}{2}+
\frac{\omega^{2}\vec{S}\cdot\vec{L}}{2mc^{2}}+\frac{\omega^{2}\hbar^{2}
P^{2}}{16m^{3}c^{4}} .
\end{eqnarray}
We then verify that from the Taylor expansion emerges the
classical harmonic oscillator Hamiltonian $H_{0}$ and the
spin-orbit coupling $V_{oso}$ naturally. And the last term is a
extra kinetic contribution to the model. Proceeding in same way
the harmonic oscillator Hamiltonian arises deformed as,
\begin{equation}
\label{hamil16}
\widehat{H}=\frac{\alpha(\omega)P^{2}}{2m}+\frac{m\omega^{2}\hat{r}}{2}+V_{hoso}\,,
\end{equation}
where the factor of deformation is, in this case, explicitly
dependent on the angular frequency $\omega$ of the system and
$V_{hoso}$ is the potential term of the harmonic oscillator that
couples to the spin-orbit term. But the kinetic energy of harmonic
oscillator is deformed by the factor $\alpha(\omega)$, which in
this case is
\begin{equation}
\label{fac3}
\alpha(\omega)=1+\frac{1}{18}\Big(\frac{E_{g}}{E_{r}}\Big)^{2}\,.
\end{equation}
Observe that the above expression is very similar to the
expression \eqref{fac2}. In $3D$ the harmonic oscillator has a
ground state energy which is dependent of the angular frequency,
\begin{equation}
E_{g}(\omega)=\frac{3\hbar \omega}{2},
\end{equation}
which in the case where we have low values of angular frequency we
have that $\frac{E_{g}}{E_{r}}<<1$ and so $\alpha(\omega)\approx
1$ which implies that we reach the usual classical harmonic
oscillator.   However, when $\omega$ is large the rate
$\frac{E_{g}}{E_{r}}$ rapidly increases which causes a deformation
on the kinetic energy of the system. Such effect could be observed
in high energy physics where the frequency $\omega$ is high too.
There exists a possibility that these effects could be observed at
nuclear or particles physics. In these cases the deformation
factor (\ref{fac3}) increases quadratically in $\omega$,
\begin{equation}
\label{fac4}
\alpha(\omega)=1+\beta\omega^{2}\,,
\end{equation}
where the constant $\beta$ depends on the mass of the system. In
the confined electron case on the noncommutative domain, the
constant $\beta=2.0\times10^{-43}s^{-2}$ and so the noncommutative
effects increases quadratically with frequency of system, as shows
the table below,
\begin{center}
\begin{tabular}{|r|c|c|}
\cline{1-3} \hline $\hspace{0.5cm}\alpha(\omega)\hspace{0.5cm}$ & $\hspace{0.5cm}\omega\,\,\,\,(Hz)\hspace{0.5cm}$ &$\hspace{0.5cm}\hspace{0.5cm}$ \\
\cline{1-3} \hline $\hspace{0.5cm}1.00\hspace{0.5cm}$ &$\hspace{0.5cm}\leq 1.0\times10^{18}\hspace{0.5cm}$ &$\hspace{0.5cm}\leq \hbox{ultraviolet frequencies}\hspace{0.5cm}$ \\
\cline{1-3} \hline $\hspace{0.5cm}1.05\hspace{0.5cm}$ &$\hspace{0.5cm}5.0\times10^{20}\hspace{0.5cm}$ &$\hspace{0.5cm}\hbox{X-ray frequency}\hspace{0.5cm}$ \\
\cline{1-3} \hline $\hspace{0.5cm}1.20\hspace{0.5cm}$ &$\hspace{0.5cm}1.0\times10^{21}\hspace{0.5cm}$ &$\hspace{0.5cm}\hbox{$\gamma$-ray frequency}\hspace{0.5cm}$ \\
\cline{1-3} \hline $\hspace{0.5cm}6.00\hspace{0.5cm}$ &$\hspace{0.5cm}5.0\times10^{21}\hspace{0.5cm}$ &$\hspace{0.5cm}\hbox{$\gamma$-ray frequency}\hspace{0.5cm}$ \\
\cline{1-3} \hline $\hspace{0.5cm}20.00\hspace{0.5cm}$ &$\hspace{0.5cm}1.0\times10^{22}\hspace{0.5cm}$ &$\hspace{0.5cm}\hbox{highest frequencies}\hspace{0.5cm}$ \\
\cline{1-3} \hline
\end{tabular} \, ,
\end{center}
where we can see that to the frequencies $\leq 10^{18}Hz$ (limit
of ultraviolet frequency) the kinetic energy is unchanged. In this
case Lorentz violation is no longer observed. However, a small
fluctuation can be noted for $5.0\times10^{20}Hz$ on the kinetic
energy of particle. Such effects can be subtly noted at nuclear
physics and at high energy physics. For highest frequencies the
factor $\alpha(\omega)$ increases rapidly up to the system reaches
an unstable state. On the other hand, for a proton confined in a
noncommutative domain we find that $\beta=5.7\times10^{-50}s^{-2}$
and so the above table comes up changed. Actually stable effects
on noncommutative space are expected for confined objects with
small mass (small $\beta$ parameter) and high energy.

\section{Quantization}

Now we are able to construct the Fock space starting off from the
Hamiltonian (\ref{hamil16}) in the absence of the spin-orbit term.
So we can write a particular Fock space which is modified by
noncommutative deformations as follows:
\begin{eqnarray}
a_{i}=\sqrt{\frac{m\omega}{2\hbar}}\,r_{i}+i\sqrt{\frac{\alpha(\omega)}{2m\hbar\,\omega}}\,P_{i}
\;\;\;\; \mbox{and} \;\;\;\;
a^{\dag}_{i}=\sqrt{\frac{m\omega}{2\hbar}}\,r_{i}-i\sqrt{\frac{\alpha(\omega)}{2m\hbar\,\omega}}\,P_{i}\,,
\end{eqnarray}
where $r_{i}=\big(x,y,z\big)$ is the ordinary displacement vector.
The particle number operator $\widehat{N}$ in noncommutative
geometry is given then by,
\begin{equation}
\widehat{N}=a^{\dag}_{i}a_{i}=
\frac{1}{\hbar\omega}\left(\frac{\alpha(\omega)P^2}{2m}+\frac{m\omega^{2}r}{2}\right)
-\frac{3}{2}\sqrt{\alpha(\omega)},
\end{equation}
and so the Hamiltonian (\ref{hamil16}) can be rewritten as,
\begin{equation}
\widehat{H}=\left(\widehat{N}+\frac{3}{2}\sqrt{\alpha(\omega)}\right)\hbar\omega\,
\end{equation}
Hence the ground state $\widehat{E}_{g}$ of noncommutative
Hamiltonian $\widehat{H}$ is correlated to the ground state of
energy of ordinary harmonic oscillator $E_{g}$ through of the
equation,
\begin{equation}
\label{fac6}
\widehat{E}_{g}=\frac{3}{2}\left[1+\frac{1}{18}\left(\frac{E_{g}}{E_{r}}\right)\right]^{\frac{1}{2}}\hbar\omega.
\end{equation}
It indicates that the system undergoes a deformation due to a
self-interaction of its own energy on a noncommutative geometry.
And such a interaction implicates in a manifest Lorentz symmetry
violating at this level. In the limit $\frac{E_{g}}{E_{r}}\approx
0$ we might obtain that $\widehat{E}_{g}\approx E_{g}$ as can be
deduced of (\ref{fac6}). We can suitably explore this violation
bearing in mind the starting point of this model. We can observe
that the equation (\ref{fac6}) shows that the energies of ordinary
harmonic oscillator becomes deformed in noncommutative geometry
(for small length scales) due the uncertainty over spacetime
operator. On the other hand, there is an iterative process in the
energy ${E}_{g}$ to reach $\widehat{E}_{g}$, therefore we can
conclude saying that any realistic noncommutative theory might be
equivalent to a truncated expansion of the original (commutative)
model. In this case, if we assume that the deformation of geometry
is due to the intrinsic spin perturbation it is possible recover
the ground state of energy in noncommutative geometry iteratively
through the ordinary ground state of energy $E_{g}$ by using the
equation (\ref{fac6}).

In order to analyze a little bit more the $\alpha$ deformation
factor we can observe that the equations (\ref{fac2}) and
(\ref{fac3}) show the same standard functional form, which is
explicitly dependent on the rate $\frac{E_{g}}{E_{r}}$. So, we can
conjecture that the maximal value for the factor of deformation
$\alpha$, taking the maximal deformation expansion reached in the
general noncommutative Hamiltonian $\widehat{H}$, is form,
\begin{equation}
\label{fac5}
\alpha(E_{g})=1+\kappa\left(\frac{E_{g}}{E_{r}}\right)^{2}\,,
\end{equation}
where $\kappa$ is a constant coefficient. The fraction
$\frac{E_{g}}{E_{r}}$ is spread out on whole model what we can
assume as a standard factor which depends on the ground state of
energy of system. An interesting point to mention is that the
general factor (\ref{fac5}) can determine the analytical form for
the ground state of energy in any system. In this case, we just
request for the noncommutative potential energy
$\widehat{V}(\hat{r})$ be well-defined.

Finally we treat the special case of heavy quarks system that
consists basically in two interacting quarks $\overline{Q}Q$
forming a quarkonium system. For a non-relativistic treatment we
can approximate the binding quark state by means of the
Schrödinger equation. Hence we can derive a noncommutative
Hamiltonian for the quarkonium system whose analytical expression,
taking the the deformation factor in its generic form
(\ref{fac6}), can be written as,
\begin{equation}
\widehat{H}=\frac{P^2}{2m}+\widehat{V}(\hat{r}),
\end{equation}
where $\widehat{V}(\hat{r})$ is the general quark interaction
potential which is given by the simple power law function in the
noncommutative version, or
%\begin{equation}
%\label{fac7}
$\widehat{V}(\hat{r})=\varphi+\lambda\hat{r}^{\nu}$,
%\end{equation}
where $\varphi$, $\lambda$ and $\nu$ are arbitrary parameters. In
same way, we might obtain the Taylor expansion of
$\widehat{V}(\hat{r})$ by assuming the equations (\ref{rcomut})
and (\ref{rspin}), which result in,
\begin{eqnarray}
\label{hamil17}
\widehat{H}=\frac{P^2}{2m}+{V}({r})+V_{so}+K_{kin}=\frac{P^2}{2m}+\varphi+\lambda r^{\nu}+
\frac{\nu\,\lambda \, r^{\nu}\vec{S}\cdot\vec{L}}{2\,m^{2}\,c^{2}\,r^2}
+\frac{\nu\,\lambda \,r^{\nu}\hbar^{2}P^{2}}{16\,m^{4}\,c^{4}\,
r^2}\; ,
\end{eqnarray}
where $K_{kin}$ is an extra kinetic contribution term to the
Hamiltonian. The above Hamiltonian $\widehat{H}$ corresponds to
the general deformed model written as,
\begin{equation}
\label{hmilt7}
\widehat{H}=\frac{\alpha(r)P^{2}}{2m}+\varphi+\lambda r^{\nu}+V_{so},
\end{equation}
where the factor of deformation $\alpha(r)$, for this case, is
given by,
\begin{equation}
\label{fac8}
\alpha(r)=1+\frac{\nu\,\lambda\,\hbar^{2}\, r^{\nu}}{8\,m^{3}\,c^{4}\,r^2}\,.
\end{equation}
Again the maximum value of $\alpha$ occurs at a particular radius
$r=a_{0}$, which, in this case denotes the fundamental radius for
the ground state of energy of the quarkonium binding system. We
can easily compute the ground state of energy of quarkonium system
plug the general factor of deformation (\ref{fac5}) in the
expression \eqref{fac8}, which is given by,
\begin{equation}
\label{fac9}
E_{g}=\eta\, \hbar\, \sqrt{\frac{\lambda \,{a_{0}}^{\nu-2}}{m}}\,,
\end{equation}
here $\eta=\sqrt{{\nu}/{8\kappa}}$ is a constant. We are able to
estimate the fundamental radius $a_{0}$ for the quarkonium system
as $a_{0}\sim\big(m\lambda\big)^{-\frac{1}{2+\nu}}$ and
substituting in (\ref{fac9}) we find that,
\begin{equation}
\label{fac10}
E_{g}\simeq\eta\hbar\frac{\left(m\lambda\right)^{\frac{2}{2+\nu}}}{m}
\end{equation}
what we can notice that it is very similar to the analytical form
for the ground state of quarkonium system which is obtained from
the noncommutative assumption on the coordinates of spacetime. It
is interesting to see that the energy in \eqref{fac10} depends
basically on the reduced mass $m$ of system as well as on the
coupling coefficient stressed by  $\lambda$ and it represents the
energy of the ordinary ground state of quarkonium system. The
equation (\ref{fac10}) fits as a good expression to the well-known
classical relation of the level spacing of energy $\Delta E$
between quarks \cite{quark1}, that depends on the reduced mass $m$
of system and of coupling strength $\lambda$, given by,
\begin{equation}
\Delta E\simeq\frac{1}{m}\big(m\lambda\big)^{\frac{2}{2+\nu}}.
\end{equation}
Therefore, strictly speaking, we can see that the parameter $\nu$
above assumes some special values when: Coulomb-like ($\nu=-1$),
simple harmonic oscillator ($\nu=2$), linear potential ($\nu=1$),
logarithm potential ($v=0$) and finally quark interaction when
($\nu=0.1$). In the special case of a quark model the equation
obtained in (\ref{fac10}) becomes consistent to the quarkonium
mass spectrum
\cite{quark1}.

\section{Conclusion}

In this work we deal with the spin structure of a charged particle
as a possible deformation to the noncommutative geometry \cite{madore}. We have
reassessed the electrostatics interaction theory where we have
embedded it in a deformed geometry and have considered that
quantum fluctuations of the spacetime implies in noncommutative
effects at very small lengths. In this sense, we have verified
that the conventional Coulomb force is modified at the length
scale of $r\leq 100fm$, and we have obtained that the effective
electric force fades away with the decreasing linearly with the
distance $r$ in such way that it vanishes in the limit
$r\rightarrow 0$. Furthermore we may infer that a screening effect
of the spacetime implies in the modified electrostatics force. We
have reasons to believe that, in this scale, the source of the
deformation of electrostatic force can be associated to quantum
basics of the discretization of spacetime.  On the other hand, for
length scales where $r> 100fm$, we have obtained that the
electrostatic force converges to the conventional Coulomb force.
In this case, we figure out that the noncommutative parameter
$\theta$ may represent a macroscopic quantity of discrete
spacetime that is associated to the open area spanned by the
vectors (or closed surface) while it becomes a nonlocal one.
Hence, as a natural consequence, the physical particles can be
viewed as nonlocal objects immersed into this deformed geometry,
moreover we have obtained that the self-energy of the electron
becomes finite when we assume the limit $r\rightarrow 0$ in this
geometry. However, we note that the self-energy turns out to be
dependent on an intrinsic deformation of the space $\rho$ which
has a quantum essence. We also verify that if $\rho\rightarrow 0$
(classical limit of spacetime) the self-energy returns to be
divergent. In this sense, it is reasonable claim that spacetime
could show an internal quantum structure whenever it is beyond the
conventional geometry.

A very direct and simple way to induce a deformation on the
spacetime is to consider that noncommutative parameter $\theta$ to
be proportional to the intrinsic spin structure \cite{madore}. Then, we can
conjectured that the area (or $\theta$) is proportional to the
modulus of the spin vector. So we could claim that the spin
structure deforms the spacetime in the quantum level what implies
in several effects in conventional Quantum Mechanics. Taking this
point as our strategy, we have obtained local spin-orbit couplings
as the linear contribution to the potential energy. In the
classical scenario we can verify that a usual potential can be
generally written down in terms of noncommutative coordinates, and
consequently the potential-like $\widehat{V}(\hat{r})$ includes a
spin-orbit coupling effects, which is in fact an angular momenta
effect due to the deformation of the trajectory. In this work, we
have obtained behaviors in a very good agreement with conventional
spin-orbit terms of Quantum Mechanics, at least as the linear
approximation. So we could say that spin-orbit effects could be
realized as noncommutative effects of spacetime due to the spin
structure.

On the other hand we have also obtained that the second-order
contribution of potential energy expansion includes an extra
kinetic term in the Hamiltonian in consideration. We verify that
this term is relevant to the dynamical structure of confined
particles submitted to the general potential energy $V(r)$ and the
kinetic term becomes a deformed one by an standard factor $\alpha$
which is dependent on the ratio between the ground state of energy
of confined system and rest energy of particle. Dealing with a
simple model which has this deformation factor $\alpha$ in
conventional physical systems we obtain, with a very good
estimative, the ground state expression as well as others
interesting aspects to the quarkonium system for instance.
Furthermore we notice that when $\alpha\approx 1$ (no kinetic
deformation) we reach easily an Hydrogen Hamiltonian system
emphasizing the Quantum Mechanics connection. We conclude that
this simple model can easily inform about the energy ground state
of complex physical systems when assuming the noncommutative
potential energy with a spin structure. Moreover we can infer that
the standard factor increases for high energy confined systems
(for instance, the quarkonium systems) where indeed such  effects
can be involved with possible Lorentz symmetry breaking.

We could comment that in order to realize the characteristic spin,
or actually its angular momentum, of a model we have to assume
that the moving charged particle is embedded into a region
fulfilled by a magnetic background field and so simulating a local
Zeeman effect on the states of the charged particle. So the
noncommutativity property is play by the background magnetic field
which is detected by the spin structure of the dynamics of the
model. Moreover we can easily infer that the atomic bound state
can be a particular case where an electron is embedded on the
nuclear electric background field.  A deeper quantum analysis will
be the object of study in a forthcoming work.

\section{Acknowledgments}
\noindent
The one of the authors, LPC, would like to thank to M.T.D. Orlando
and A. Biondo for the kind hospitality in Physics Department of
the Universidade Federal do Esp\'\i rito Santo (UFES)- Brasil. WCS
is grateful to CAPES and CNPq for the financial support.

\end{document}